\documentclass[prb,aps,twocolumn]{revtex4}
\usepackage{graphicx,color,latexsym}
\usepackage{dcolumn}
\usepackage{amsmath,amssymb,epsf,bm}
\begin{document}
\title{Antiferromagnetic Josephson junction: nonreciprocity and sublattice selective transport of Cooper triplets}
\author{A.~G. Mal'shukov}
\affiliation{Institute of Spectroscopy, Russian Academy of Sciences, Troitsk, Moscow, 108840, Russia}

\begin{abstract}
The  Josephson junction under consideration is composed of two  s-wave superconducting contacts deposited on the top of a  two-dimensional antiferromagnet (AF). Triplet Cooper correlations in AF are provided by thin ferromagnetic spacers between AF and superconducting contacts. The problem is considered in the regime of a weak tunneling of electrons between these contacts and AF.  The transport of  electron Cooper's pairs under the stationary phase bias in the disordered AF is treated within the formalism of equilibrium Green functions and the Born approximation for electrons which are scattered by nonmagnetic impurities. The system of diffusion equations is derived for three triplet components of Cooper-correlated electron pairs. These spin projections are coupled to each other due to interplay of the spin-orbit interaction and AF order. Moreover, the diffusion equations are sublattice-dependent. It is shown that such a staggered behavior will lead to the nonreciprocal Josephson current, if the tunneling of electrons between  contacts and AF is different for different sublattices.

\end{abstract}
\maketitle

\section{Introduction}

The physics of systems which combine superconductivity and magnetism was intensively studied during last decades. Although both of these phenomena tend to compete and suppress each other, being combined in one system, for instance in heterostructures comprising magnetic and superconducting materials, they may reveal strikingly new effects and realize new functionalities.
Most works in this field are focused  on electronic transport in superconductor-ferromagnetic systems which were discussed in several reviews \cite{Linder1,Buzdin,Bergeret,Robinson}. The interplay of superconductivity and ferromagnetism can also reveal unusual topological properties of superconductor-ferromagnetic heterostructures. \cite{Qi}.

Superconductor-antiferromagnet (AF) junctions attracted initially not so broad interest. On the other hand, recent results on AF have demonstrated that, despite of their magnetic "neutrality", these materials can be successfully incorporated into various spintronic applications due to their outstanding spin transport characteristics \cite{Baltz,Gomonay,Yan,Wadley}. Therefore, interesting results may also be expected from studies of superconductor-AF heterostructures. Theoretical and experimental works in this field were mainly focused on studying AF-S bilayers and  multilayers. In these heterostructures both materials may affect physical properties of each other. \cite{Krivoruchko,Linder2021,Wu,Hubener,Bobkov2,Bobkov1,Brataas1} For example, a contact with AF may lead to suppressing of the superconducting  critical temperature of an S layer.  At the same time,  triplet superconducting correlations  can be induced in superconducting and AF layers by the N\'{e}el magnetic order. Also, new bound Andreev states were predicted at an AF-S interface \cite{Bobkova,Barash2005}. In some works Josephson junctions (JJ) were considered, whose weak links are represented by AF. Measurements of the  critical current in S-AF-S junctions revealed its fast reduction at thicknesses of AF layers exceeding several nanometers.\cite{Bell,Weides} This result is corroborated by measurements of the Cooper pair's penetration depth into an AF near its interface with a superconductor. \cite{Hubener} In contrast, the long-range coherence length was reported in weak links containing copper oxide antiferromagnetic layers. \cite{Zaitsev,Zaitsev2} Such a phenomenon was theoretically explained \cite{Zaitsev} within a model of a JJ which is formed of antiferromagnetically ordered bulk ferromagnetic layers. A similar model was also considered  in Ref.[\onlinecite{Gorkov}]. A planar version of such sort of junctions, with ferromagnetic layers replaced by atomic chains, have been studied in Refs.[\onlinecite{Andersen,Falch}]. In almost all quoted above theoretical works on JJ authors assumed that normal metallic interlayers are free of defects, so that the transport of electrons within the layers is ballistic. An exception is Ref.[\onlinecite{Zaitsev}], where the bulk disordered ferromagnetic layers were treated within the semiclassical theory, which was previously applied for ferromagnetic JJ. \cite{Buzdin,Bergeret}

At the root of most important physical phenomena in AF-S heterostructures is the superconducting proximity effect which results in  superconducting correlations of electrons in  AF.  The early theory of this effect in disordered metallic AF layers of an AF-S superlattice has been presented in Ref.[\onlinecite{Krivoruchko}], where antiferromagnetism was considered as a spin density wave of itinerant electrons. In this work the theory of semiclassical  Green functions \cite{Eilenberger,Larkin semiclass} was applied for studying the superconducting proximity effect in such sort of AF. This theory predicted a short coherence length of superconducting correlations in disordered AF, very similar to that in ferromagnets. A generalization of the semiclassical  Green functions formalism  on the bipartite crystalline AF has been carried out in Refs.[\onlinecite{Bobkov2,Bobkov1,Brataas1,Fynn}] for  S-AF bilayers. It has been shown that the s-wave Cooper pairing in the  superconductor produces the short-range proximity effect in the AF layer. At the strong enough exchange interaction between itinerant electrons and localized spins superconducting correlations may extend in the AF over lengths as small as the electron mean free path. \cite{Brataas1} In contrast, the long-range effect is possible for triplet Cooper correlations whose spin is perpendicular to the N\'{e}el vector of the AF. \cite{Brataas1}

Although some important details of the superconducting proximity effect in AF-S heterostructures are now clarified, the question remains unanswered, as how a disordered JJ with an AF weak link will operate. In particular, it is interesting to find out the role of the long-range proximity effect, in particular, if different odd-triplet components of the electron Cooper pair correlations are mutually entangled due to the N\'{e}el order and the spin orbit coupling of electrons. In this work a planar JJ will be considered whose weak link is represented by a two-dimensional (2D) disordered metallic AF. This JJ is phase biased, while s-wave bulk superconductors make contacts with the AF film by means of thin ferromagnetic layers, as shown in Fig.1. The ferromagnetic layers provide an admixture of long-range Cooper triplets in AF. A regime of the weak tunnel coupling through AF-S contacts will be assumed. Within this approximation the calculation of the Josephson current was carried out without using the semiclassical Green functions formalism. So, diffusion equations were obtained for mutually coupled vector components of triplet correlation functions. On the basis of these equations the dependence of the  Josephson current on orientations of the magnetization in ferromagnetic layers was calculated. The diffusion equations turned out to be sublattice selective, depending on magnetization directions of contacts. A nontrivial consequence of this sublattice dependence is that the Josephson current may become nonreciprocal, if electron's tunneling amplitudes are different for different sublattices in one, or both S-AF contacts. In this case the current does not reverse its direction, if the phase difference between superconducting contacts changes  the sign. Such a junction plays the  role of a $\varphi_0$-junction with $\varphi_0=\pi/2$ when the singlet supercurrent is suppressed by disorder in a sufficiently long JJ.

The article is organized in the following way. The formalism which was employed in this work will be presented in Sec. II. In Sec. III the diffusion equation for the electron pairing function in AF is derived and its solutions are analyzed.  Results are presented and discussed in Sec. IV.

\begin{figure}[tp]
\includegraphics[width=9cm]{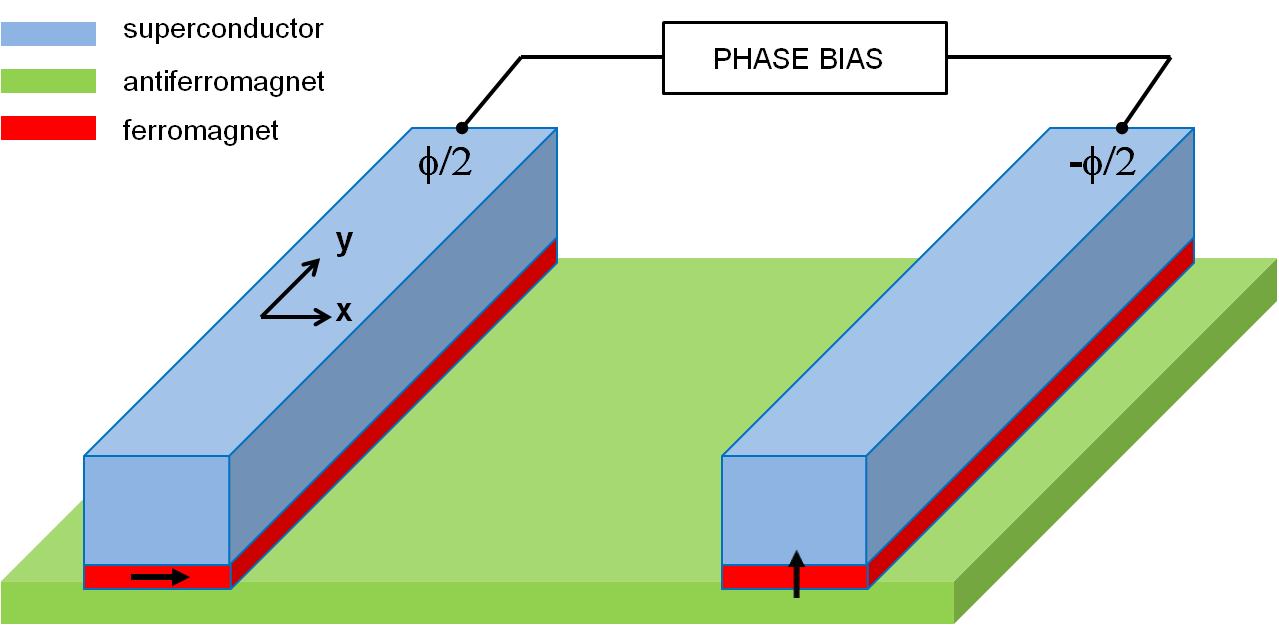}
\caption{(Color online) Two superconductors make contact via thin ferromagnetic films with a 2D antiferromagnetic metal. The phase bias provides the phase difference $\phi$ between order parameters of superconducting terminals. Magnetizations of the ferromagnetic films need not be collinear. The black arrows show one of the possible noncollinear configurations of magnetizations.} \label{fig1}
\end{figure}
\section{Formalism}
\subsection{Tunneling Hamiltonian}
In the framework of the perturbation theory \cite{Aslamazov} the Josephson current is determined by the fourth order perturbation theory with respect to the tunneling amplitude between an AF film and bulk  superconducting s-wave terminals. The latter are considered to be sufficiently massive, that allows  to ignore influence of 2D electron gas in the AF film on superconducting properties of these contacts. The  unperturbed Hamiltonian of two-dimensional (2D) electron gas in AF is given by $H_0=\sum_{\mathbf{k}}\psi^{\dag}_{\mathbf{k}}\mathcal{H}_{0\mathbf{k}}\psi_{\mathbf{k}}$, where $\psi_{\mathbf{k}}$ are the electron field operators,  which are defined in the Nambu basis as $\psi_{\mathbf{k},\lambda}=(c_{\mathbf{k},\uparrow,\lambda},c_{\mathbf{k},\downarrow,\lambda},c^{\dag}_{\mathbf{-k},\downarrow,\lambda},
-c^{\dag}_{\mathbf{-k},\uparrow,\lambda})$, where  $\lambda=1,2$ is the sublattice index and  $\sigma$ in $c_{\mathbf{k},\sigma,\lambda}$ denotes the spin projection ($\sigma=\uparrow,\downarrow$). The one-particle Hamiltonian $\mathcal{H}_{0\mathbf{k}}$ is given by
\begin{equation}\label{H0}
\mathcal{H}_{0\mathbf{k}}=\epsilon_{\mathbf{k}}\lambda_x\rho_z+\lambda_x\rho_z\bm{\sigma}\mathbf{h_{\mathbf{k}}}+J\sigma_z\lambda_z-\mu\rho_z \,,
\end{equation}
where $\epsilon_{\mathbf{k}}=-t(\cos ak_x+\cos ak_y)$ is the electronic band energy, with $t$ denoting the electron's hopping constant  between near neighbors of a bipartite square lattice, whose lattice constant is $a$. The second term represents the Rashba interaction, where  $h_{\mathbf{k}}^x=-\alpha_R\sin ak_y$ and $h_{\mathbf{k}}^y=\alpha_R\sin ak_x$.
In Eq.(\ref{H0}) $\mu$ is the  chemical potential and $J\sigma_z\lambda_z$ is the exchange interaction of electrons with the staggered N\'{e}el magnetic order, where $\bm{\sigma}=(\sigma_x,\sigma_y,\sigma_z)$ is the vector of Pauli matrices and $\lambda_i$  $(i=x,y,z)$ are Pauli matrices acting in the sublattice space, while matrices $\rho_i$  ($i=x,y,z$) operate in the Nambu space. The exchange interaction of electrons with the checkerboard antiferromagnetic order in $\mathcal{H}_{0\mathbf{k}}$ gives rise to the energy gap $2J$ just in the middle of the tight-binding electronic band. In turn, the Rashba coupling results in a spin splitting of the electron energy. The chemical potential will be counted from the middle of the band. The considered parameter's range is such that $\mu \gtrsim J$ and $\alpha_R \ll \mu$. Besides Eq.(\ref{H0}), the total Hamiltonian includes also the potential $V$ which is associated with disorder. Therefore, $H=H_0 + V$. The disorder will be modeled by random energies which should be added to the cite energy. It is assumed that these energies fluctuate independently on each cite.

We assume a weak coupling between AF and superconducting contacts. The corresponding tunneling Hamiltonians $H_{L}$ and  $H_{R}$  for the left and right contacts can be written in the form
\begin{equation}\label{Hint}
H_{L(R)}=\frac{1}{\sqrt{V_{L(R)}}}\sum_{\mathbf{k},\mathbf{k}^{\prime}}(\psi^{\dag}_{\sigma\lambda\mathbf{k}}t^{\lambda}_{L(R)\mathbf{k},\mathbf{k}^{\prime}}
\rho_z\psi^{s}_{L(R)\sigma\mathbf{k}^{\prime}}+ h.c.)\,,
\end{equation}
where $\psi^{s}_{L(R)\mathbf{k}^{\prime}}$ and $V_{L(R)}$ are the electron's field operators  and volumes of left and right superconducting contacts, respectively. Note, that tunneling parameters $t^{\lambda}_{L(R)\mathbf{k},\mathbf{k}^{\prime}}$ may in general depend on the sublattice index $\lambda$. Each superconducting contact gives rise to the electron's self-energy $\Sigma_{L(R)}$, which may be written in the form
\begin{eqnarray}\label{Sigma}
&&\Sigma^{\lambda\lambda^{\prime}}_{L(R)\mathbf{k},\mathbf{p}}(\omega_n)=\frac{1}{V_{L(R)}}\sum_{\mathbf{k}^{\prime}}t^{\lambda}_{L(R)\mathbf{k},\mathbf{k}^{\prime}}
t^{\lambda^{\prime}}_{L(R)\mathbf{k}^{\prime},\mathbf{p}}\times\nonumber \\
&&\rho_zG^{s}_{L(R)\mathbf{k}^{\prime}}(\omega_n)\rho_z\,,
\end{eqnarray}
where $G^{s}_{L(R)\mathbf{k}}(\omega_n)$  are impurity averaged Matsubara Green's functions of electrons in the left and right superconducting terminals, with frequencies $\omega_n=\pi (2n+1)T$ ($n$=0,1,2...) where the Boltzman  constant is set to unity. Since, as shown in Fig.1, these contacts have thin ferromagnetic interlayers, the functions $G^{s}_{L(R)\mathbf{k}}(\omega_n)$ depend on electron spins. Therefore, they are represented by 4$\times$4 matrices operating in the spin and Nambu spaces. Within the perturbation theory only nondiagonal matrix elements with respect to Nambu variables are important in Eq.(\ref{Sigma}). Therefore, we retain in this equation only these nondiagonal parts of $G^{s}_{L(R)\mathbf{k}}$. They are anomalous Green functions $F^{s}_{L(R)\mathbf{k}}$ which are associated with respective superconducting order parameters $\Delta_{L(R)}$. It is easy to see that in Eq.(\ref{Sigma}) $\rho_zF^{s}_{L(R)}\rho_z=-F^{s}_{L(R)}$, because $F^{s}_{L(R)\mathbf{k}}(\omega_n)$ has no diagonal elements in the Nambu space. Further, Eq.(\ref{Sigma}) can be considerably simplified by assuming that the S-AF interface is disordered. So, $t^{\lambda}_{L(R)\mathbf{k}\mathbf{k}^{\prime}}$  becomes isotropic with respect to directions of vectors $\mathbf{k}$ and $\mathbf{k}^{\prime}$, while the vector's magnitudes stay fixed near the respective Fermi surfaces of S and AF. Therefore, the disorder average  of transmission amplitudes in Eq.(\ref{Sigma}) may be represented as $\langle t^{\lambda}_{L(R)\mathbf{k},\mathbf{k}^{\prime}}t^{\lambda^{\prime}}_{L(R)\mathbf{k}^{\prime},\mathbf{p}}\rangle=|t^{\lambda}_{L(R)}||t^{\lambda^{\prime}}_{L(R)}|\langle\exp i(\theta_{\lambda}-\theta_{\lambda^{\prime}})\rangle$, where $\theta_{\lambda}$ and $\theta_{\lambda^{\prime}}$ are phases of the transmission amplitudes. At sufficiently strong disorder the phase factor at $\lambda\neq\lambda^{\prime}$ can turn to zero due to the averaging. Below, for simplicity, only diagonal terms of the self-energy will be taken into account. As a result, by taking into account the isotropy of transmission amplitudes with respect to the wave vector,  Eq.(\ref{Sigma}) can be transformed into
\begin{equation}\label{Sigma2}
\Sigma^{\lambda\lambda^{\prime}}_{L(R)\mathbf{k},\mathbf{p}}(\omega_n)=-\delta_{\lambda\lambda^{\prime}}\frac{|t^{\lambda}_{L(R)}|^2}{V_{L(R)}}\sum_{\mathbf{k}^{\prime}}
F^{s}_{L(R)\mathbf{k}^{\prime}}(\omega_n)\,,
\end{equation}

The electric current through the junction may be expressed \cite{Aslamazov} by calculating the time derivative of the total charge in one of the superconducting contacts. The commutator of this charge in the left contact with tunneling Hamiltonian Eq.(\ref{Hint}) gives the current operator  in the form
\begin{equation}\label{jhat}
\hat{j}=\frac{ie}{\sqrt{V_{L(R)}}}\sum_{\mathbf{k},\mathbf{k}^{\prime}\lambda}(\psi^{\dag}_{\sigma\lambda\mathbf{k}}t^{\lambda}_{L(R)\mathbf{k},\mathbf{k}^{\prime}}\psi^{s}_{L
\sigma\mathbf{k}^{\prime}}- h.c.)\,,
\end{equation}
As follows from  Eq.(\ref{jhat}), the current operator is proportional  to the tunneling amplitude. Therefore, in order to obtain the Josephson current, it is necessary to calculate the thermodynamic average of  Eq.(\ref{jhat}), up to the fourth order with respect to $t_{L(R)}$.  Within the Matsubara formalism \cite{AGD}, by considering Eq.(\ref{Hint}) as a perturbation, \cite{Aslamazov} we obtain from Eq.(\ref{jhat}) the Josephson current in the form
\begin{eqnarray}\label{j}
&&j=ieT\sum_{\omega_n}\sum_{\mathbf{p},\mathbf{p}^{\prime},\mathbf{k}^{\prime}\mathbf{k}}\mathrm{Tr}\langle\Sigma_{L\mathbf{k},\mathbf{p}}(\omega_n)\rho_z
G_{\mathbf{p}\mathbf{p}^{\prime}}(\omega_n)\times\nonumber \\
&&\Sigma_{R\mathbf{p}^{\prime}\mathbf{k}^{\prime}}(\omega_n)G_{\mathbf{k}^{\prime}\mathbf{k}}(\omega_n)
\rangle_{av}\,,
\end{eqnarray}
where $\langle...\rangle_{av}$ denotes the averaging of the Green's functions product over disorder in the antiferromagnetic metal. These functions are 2$\times$2 matrices in spin, sublattice and Nambu spaces. They depend on two wave vectors, as long as they are not averaged over impurities. At the same time, the averaged functions $G_{\mathbf{p}}(\omega)$ are homogeneous in space. Therefore, they depend only on a single wave vector, so that  $\langle G_{\mathbf{p}\mathbf{p}^{\prime}}(\omega_n)\rangle_{av} = G_{\mathbf{p}}(\omega_n)\delta_{\mathbf{p}\mathbf{p}^{\prime}}$

\subsection{Impurity averaging}

The current in Eq.(\ref{j}) can be expressed in terms of the Cooper pair correlation function, which  is given by
\begin{equation}\label{K}
K_{\mathbf{p}\mathbf{p}^{\prime}\mathbf{k}^{\prime}\mathbf{k}}(\omega_n)=
\langle G^{\rho}_{\mathbf{p}\mathbf{p}^{\prime}}(\omega_n)\otimes G^{\rho^{\prime}}_{\mathbf{k}^{\prime}\mathbf{k}}(\omega_n)\rangle_{av}\,,
\end{equation}
where $\rho$ and $\rho^{\prime}$ are Nambu variables which take on the values $\pm1$. Since AF is a normal metal, its Green functions are diagonal in the Nambu space. Therefore, they may be characterized by corresponding eigenvalues of $\rho_z$, so that at $\rho=\pm1$ the Green's functions represent  the particle and hole sectors of the Nambu space, respectively. For the Cooper pair correlation function in Eq.(\ref{K}) these eigenvalues  are such, that $\rho\neq \rho^{\prime}$. Hence, the correlation function in Eq.(\ref{K}) represents a correlated pair of an electron and an Andreev-reflected hole. The disorder-averaged Green's functions are the main entries in $K$. They satisfy the Dyson equation
\begin{equation}\label{Dyson}
(i\omega_n-H_{0\mathbf{k}}-\Sigma_{imp}(\omega_n))G_{\mathbf{k}}(\omega_n)=1\,,
\end{equation}
where  $\Sigma_{imp}$ represents the self-energy, which is associated with the disorder. The latter is given  by random uncorrelated shifts $u_i$ of  site energies. Their pair correlation function is $\overline{u_iu_j}=u^2\delta_{ij}$.  By taking into account the latter equation, within the Born approximation the self-energy can be written as \cite{Rammer}
\begin{equation}\label{sigmaimp}
\Sigma^{\lambda\lambda^{\prime}}_{imp}(\omega)=\delta_{\lambda\lambda^{\prime}}\frac{u^2}{N_c}\sum_{\mathbf{k}}G^{\lambda\lambda}_{\mathbf{k}}(\omega_n)\,,
\end{equation}
where the superscripts $\lambda$ and $\lambda^{\prime}$ denote sublattice elements of the matrix $\Sigma_{imp}$, while $N_c$ is the number of  unit cells in the AF crystal. As seen from Eq.(\ref{sigmaimp}), this matrix is diagonal. It is important to calculate the imaginary part of Eq.(\ref{sigmaimp}), because it determines major relaxation mechanisms of  particle and spin transport. At the same time, the real part of the self-energy can be included in the band energy. It is seen from Eq.(\ref{sigmaimp}) that  $\mathrm{Im}(\Sigma_{imp})$ is proportional to $\mathrm{Im}(\sum_{\mathbf{k}}G_{\mathbf{k}})$. In the leading approximation $G_{\mathbf{k}}(\omega_n)$ may be found from Eq.(\ref{Dyson}) by neglecting there small terms, such as the Rashba coupling and $\Sigma_{imp}$. By denoting this function in the respective sector $\rho$ of the Nambu space as  $G^{\rho}_{0\mathbf{k}}(\omega_n)$, we obtain it from Eq.(\ref{Dyson}) in the form
\begin{equation}\label{G0}
G^{\rho}_{0\mathbf{k}}(\omega_n)=\frac{1}{2}\sum_{\beta=\pm1}\frac{(1+\beta \hat{P}^{\rho}_{\mathbf{k}})}{i\omega_n+\mu\rho-\beta E_{\mathbf{k}}}\,,
\end{equation}
where $E_{\mathbf{k}}=\sqrt{\epsilon^2_{\mathbf{k}}+J^2}$ and $\hat{P}^{\rho}_{\mathbf{k}}=(\epsilon_{\mathbf{k}}\rho\lambda_x+J\lambda_z\sigma_z)/E_{\mathbf{k}}$. The summation of  $G_{0\mathbf{k}}(\omega_n)$ over $\mathbf{k}$ in Eq.(\ref{sigmaimp}) gives
\begin{equation}\label{sigmaimp2}
\Sigma_{imp}(\omega_n)=i\Gamma\frac{\omega_n}{|\omega_n|} \left(1+\lambda_z\sigma_z\frac{J}{\mu}\right)\,,
\end{equation}
where  $\Gamma=\pi u^2N_{\mu}/2$ and $N_{\mu}$ is the state density at the Fermi level per one unit cell. By substituting  Eq.(\ref{sigmaimp2})  in Eq.(\ref{Dyson}) the  disorder-averaged Green's function can be expressed as
\begin{equation}\label{G}
G^{\rho}_{\mathbf{k}}(\omega_n)=\frac{1}{4}\sum_{\beta,\gamma=\pm1}\frac{(1+\beta \hat{P}^{\rho}_{\mathbf{k}})(1+\gamma \hat{R}^{\rho}_{\mathbf{k}})}{i\omega_n+i\tilde{\Gamma}(\omega_n)+\mu\rho-\beta E_{\mathbf{k}}-\gamma h_{\mathbf{k}}}\,,
\end{equation}
where $\hat{R}^{\rho}=\rho\lambda_x\bm{\sigma}\mathbf{h}_{\mathbf{k}}/  h_{\mathbf{k}}$ and $\tilde{\Gamma}(\omega_n)=(\omega_n/|\omega_n|)\Gamma(1+J^2/\mu^2)$.
\begin{figure}[tp]
\includegraphics[width=6cm]{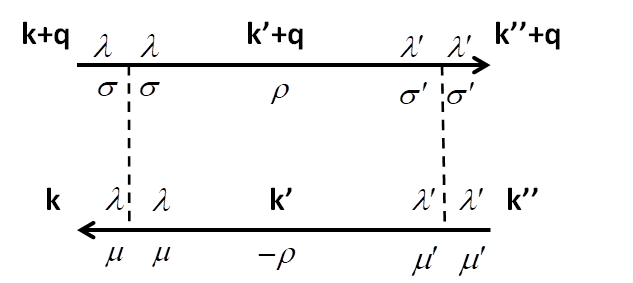}
\caption{A fragment of the ladder series. It represents the diffusion motion of electrons in a disordered antiferromagnet. The dashed lines correspond to the impurity scattering, $\lambda$ and $\lambda^{\prime}$ are sublattice indexes, while $\sigma$ and $\mu$ are spin variables. $\rho$ denotes the Nambu variable. Its opposite signs for two Green functions corresponds to propagation of a correlated pair composed of an electron and an Andreev-reflected hole.} \label{fig2}
\end{figure}
Within the semiclassical approximation the correlation function, which is given by Eq.(\ref{K}), is represented by ladder series of Feynman diagrams. These diagrams describe multiple scattering of electrons by disorder. Such a scattering results in diffusive transport of particles. Due to the averaging in Eq.(\ref{K}) the wave vectors satisfy the equations $\mathbf{p}=\mathbf{k}+\mathbf{q}, \mathbf{p}^{\prime}=\mathbf{k}^{\prime}+\mathbf{q}$. By denoting
\begin{equation}\label{D}
D_{\mathbf{q}}(\omega_n)=(u^2/2)\sum_{\mathbf{k,k^{\prime}}}K_{\mathbf{k+q},\mathbf{k}^{\prime}+\mathbf{q},\mathbf{k}^{\prime},\mathbf{k}}(\omega_n)
\end{equation}
one can see that the ladder summation results in the equation for $D_{\mathbf{q}}(\omega_n)$, which  can be written in the form
\begin{equation}\label{K2}
D_{\mathbf{q}}(\omega_n)=\Pi_{\mathbf{q}}(\omega_n)-\Pi_{\mathbf{q}}(\omega_n)D_{\mathbf{q}}(\omega_n)\,.
\end{equation}
This equation is written in a symbolic form. Spin, sublattice and Nambu variables are not shown. They are represented explicitly in a fragment of the ladder diagram in Fig.2. A single element of the expansion in powers of the disorder strength  has the form
\begin{equation}\label{Pi}
\Pi^{\rho\bar{\rho},\lambda\lambda^{\prime}}_{\mathbf{q}}(\omega_n)=\frac{u^2}{2}\sum_{\mathbf{k}}G^{\rho,\lambda\lambda^{\prime}}_{\mathbf{k}+\mathbf{q}}(\omega_n)
G^{\bar{\rho},\lambda^{\prime}\lambda}_{\mathbf{k}}(\omega_n)\,,
\end{equation}
where $\bar{\rho}=-\rho$. Accordingly, $\Pi^{\rho\bar{\rho}}$ denotes the corresponding nondiagonal element in the Nambu space. This function  represents correlated dynamics of a particle and an Andreev-reflected hole. It is convenient to introduce for $\Pi$ a new representation, instead of 2$\times2$ matrices in spin space. Let us denote
\begin{equation}\label{Piij}
\Pi^{\rho\bar{\rho},\lambda\lambda^{\prime}}_{ij,\mathbf{q}}(\omega_n)=\frac{u^2}{2}
\sum_{\mathbf{k}}\mathrm{Tr}[\sigma^iG^{\rho,\lambda\lambda^{\prime}}_{\mathbf{k}+\mathbf{q}}(\omega_n)
\sigma^jG^{\bar{\rho},\lambda^{\prime}\lambda}_{\mathbf{k}}(\omega_n)]\,,
\end{equation}
where the trace is taken over spin projections, while $ i$ and $j$ take on the values 0,x,y and z, so, the unit matrix $\sigma^0$ is added to the set of Pauli matrices. The same representation is used for $D_{\mathbf{q}}(\omega_n)$. Therefore, Eq.(\ref{K2}) should be understood as a matrix equation for $2\times2$ and $4\times4$ matrices in the sublattice and $(ij)$ spaces, respectively.

While calculating the function $\Pi$, effects were neglected which are associated with small corrections beyond the semiclassical approximation. These corrections are of the order of $q/k_F$ and $\alpha_R/t$. They lead to the coupling between spin and charge electronic degrees of freedom during  transport of electrons in normal metals, as well as to the mixing of singlet and triplet Cooper  correlations in superconducting systems. The latter  leads, for example, to the $\varphi_0$ junction effect \cite{Reinoso,Zazunov,ISHE,Liu,Yokoyama,Konschelle}, while the former results in the spin-Hall effect. When the singlet-triplet coupling is ignored, the matrix $\Pi_{ij}$  in Eq.(\ref{Piij}) decouples into the $\Pi_{00}$ singlet and nine $\Pi_{ij}$ triplets, where $i$ and $j$ take on the values $x,y$ and $z$. The dependence of $\Pi_{\mathbf{q}}(\omega_n)$ on the wave vector $\mathbf{q}$ gives rise to spatial variations of the superconducting proximity effect throughout the junction. The length of these  variations is assumed to be much larger than the mean free path of electrons. Therefore,  $\Pi_{\mathbf{q}}(\omega_n)$ will be expanded up to $q^2$ terms. Moreover, as long as Rashba coupling $\alpha_R\ll \Gamma$, this correlation function will also be  expanded up to $\alpha_R^2/\Gamma^2$ terms. Matrix elements of $\Pi_{\mathbf{q}}(\omega_n)$ are calculated in Appendix A. They will be used for the analysis of diffusion equations for the pair  correlation function $D_{\mathbf{q}}(\omega_n)$.

\subsection{Diffusion equations}

The diffusion equation for the decoupled singlet $D_{00\mathbf{q}}(\omega_n)$ is readily obtained from Eqs.(\ref{K2}) and (\ref{Pi00}). Since the matrix $\Pi$ in Eq.(\ref{Pi00}) does not depend on $\lambda, \lambda^{\prime}$ and $\rho$, the same is also valid for $D^{\rho\bar{\rho},\lambda\lambda^{\prime}}_{00,\mathbf{q}}(\omega_n)$. Therefore, by skipping sublattice and Nambu variables Eq.(\ref{K2}) can be written in the form
\begin{equation}\label{D00}
\left(2|\omega_n|+c_0Dq^2+4\Gamma\xi^2\right)D_{00,\mathbf{q}}(\omega_n)=-\Gamma(1-\xi^2)\,,
\end{equation}
where $\xi=J/\mu$, $c_0=(1-\xi^2)/(1+\xi^2)^2$  and  $D=v^2\tau_{\mathrm{imp}}/2$ is the diffusion constant. In the latter equation $\tau_{\mathrm{imp}}=1/2\Gamma$ is the mean elastic scattering time of electrons.  The first two terms in Eq.(\ref{D00}) describe the diffusion of Cooper pairs in AF. At the same time, the third term results in decreasing in AF of superconducting correlations which are induced by superconducting contacts. As seen from Eq.(\ref{D00}), at $\xi \simeq 1$  these correlations penetrate into AF over a short distance, of the order of $\sqrt{D/\Gamma}$, which is comparable to the mean free path of electrons. This result coincides with the conclusion \cite{Brataas1,Bobkov1} about  a destructive influence of nonmagnetic impurities on the superconducting proximity effect in AF. On the other hand, the long-range superconducting proximity effect may be reached at $\xi\ll 1$. This means that within the considered tight binding model the Fermi level must be far enough from the van Hove singularity in the middle of the electron energy band.

The behavior of triplet components of the function $D_{ij\mathbf{q}}(\omega_n)$ is very distinct from the singlet one. At the finite Rashba SOC these components are entangled with each other. Therefore, there is a set of three equations for $xj$, $yj$ and $zj$ matrix elements, at fixed $j\in(x,y,z)$. Besides, due to nondiagonal sublattice elements of the matrix $\Pi^{\lambda\lambda^{\prime}}_{ij\mathbf{q}}(\omega_n)$ the total number of equations is doubled. However, the $(x,y)$ and $z$ subspaces are coupled only due to $\Pi_{iz}$ and  $\Pi_{zi}$ matrix elements of the function $\Pi$ in Eq.(\ref{K2}), where $i\in(x,y)$. At the same time, according to Eq.(\ref{Pixz}) and Eq.(\ref{Piyz}), at $vqh_{\mathbf{k}}\ll \Gamma^2$ these functions are small. Therefore, one can first calculate the decoupled correlators $D_{ij}$ and $D_{zz}$, where  $ (i,j)\in(x,y)$, and then take into account their coupling  $\Pi_{iz}$ perturbatively. Within this approximation $D_{zz}$ can be obtained by retaining in Eq.(\ref{K2}) only $\Pi_{zz}$, and $D_{zz}$ terms. At $\xi \sim 1$, by neglecting in Eq.(\ref{Pizz}) small terms, which are proportional to $q^2$ and $h_{\mathbf{k}}^2$, we arrive at
\begin{equation}\label{Dzz2}
D^{\rho\bar{\rho},\lambda\lambda^{\prime}}_{zz,\mathbf{q}}(\omega_n)=-\frac{(1-\xi^2)}{4\xi^2}\,.
\end{equation}
At $\xi\ll 1$ it is necessary to take into account these omitted terms. So, in the leading approximation Eq.(\ref{K2}) gives
\begin{equation}\label{Dzz3}
(2|\omega_n|+Dq^2+\frac{2\langle h^2_{\mathbf{k}}\rangle}{\Gamma}+4\Gamma\xi^2)D^{\rho\bar{\rho},\lambda\lambda^{\prime}}_{zz,\mathbf{q}}(\omega_n)=-\Gamma \,.
\end{equation}
As seen from Eq.(\ref{Dzz2}) and Eq.(\ref{Dzz3}) the function $D^{\bar{\rho}\rho,\lambda\lambda^{\prime}}_{zz}$ does not depend on sublattice and Nambu indexes. Therefore we set $D^{\bar{\rho}\rho,\lambda\lambda^{\prime}}_{zz}\equiv D_{zz}$

To find diffusion equations for other elements of the matrix $D_{ij}$ let us consider a junction which is uniform in the $y$-direction and, accordingly, set $q_y=0$ in Eq.(\ref{Pixx}-\ref{Piyz}).  We are going to write these equations for the pair correlation function $D^{1-1}$ in the Nambu space. In turn, the function $D^{-11}$ can be expressed according to $D^{-11,\lambda\lambda^{\prime}}_{ij,\mathbf{q}}=D^{1-1,\lambda^{\prime}\lambda}_{ji,-\mathbf{q}}$ $((i,j)\in (x,y,
z))$, as shown in Appendix A. Therefore, for brevity the Nambu superscripts have suppressed. In diffusion equations, within the employed tight-binding model, the averages over the Fermi surface which have the form $\langle h_{\mathbf{k}}^2\rangle$ and $\langle h^y_{\mathbf{k}}v_x\rangle$ can be expressed via $\langle v^2\rangle$ by using the expressions for the Rashba SOC and the relation $v_i=\nabla_{k_i}\epsilon_{\mathbf{k}}=at\sin ak_i$, where $i\in(x,y)$. So, it is easy  to obtain $\langle h_{\mathbf{k}}^2\rangle=\kappa^2\langle v^2\rangle$ and $\langle h^y_{\mathbf{k}}v_x\rangle=\kappa\langle v^2\rangle/2$, where $\kappa=\alpha_R/at$. With this parametrization Eq.(\ref{K4}), Eq.(\ref{M2}),  Eq.(\ref{MDiz}) and Eq.(\ref{Iiz}) give
\begin{eqnarray}\label{D11}
&&(2|\omega_n|+c_1Dq_x^2+4c_1\kappa^2D)D^{1\lambda}_{ij}=I^{1\lambda}_{ij}\,,\nonumber\\
&&(2|\omega_n|+c_1Dq_x^2+4c_1\kappa^2D)D^{11}_{yz}=c_2\xi \kappa Dq_x (2D_{zz}-1)\,,\nonumber\\
&&(2|\omega_n|+c_1Dq_x^2+4c_1\kappa^2D)D^{11}_{xz}=-ic_2\kappa D q_x (2D_{zz}-1)\,,\nonumber \\
&&D^{12}_{iz}=D^{11}_{iz}\,,\,D^{12}_{zi}=-D^{12}_{iz}\,,
\end{eqnarray}
where  $I^{1\lambda}_{ij}$ is given by Eq.(\ref{I}), $\lambda\in(1,2)$, $(i,j)\in(x,y)$, the coefficient $c_2=2\sqrt{1-\xi^2}/(1+\xi^2)^2$ and $D_{zz}$ is expressed  by Eq.(\ref{Dzz2}) at $\xi \sim 1$. The functions $D^{22}$ and $D^{21}$ can be obtained from $D^{11}$ and $D^{12}$ by the substitution $1\rightleftarrows2$ and $\xi\rightarrow -\xi$. Other components of the matrix $D$ can be found via the  Hermitian conjugation, by taking into account that $D^{\lambda\lambda^{\prime}}_{ij}=D^{\lambda^{\prime}\lambda *}_{ji}$.

In the absence of antiferromagnetism at $\xi=0$ Eq.(\ref{K2}) reduces to the known diffusion equation for the proximity induced superconducting triplet correlations in a Rashba coupled diffusive system \cite{Malsh}. This equation contains the D'yakonov-Perel \cite{DP} spin relaxation, which is represented by the third term in the left-hand side of Eq.(\ref{D11}) and is  proportional to $\kappa^2$. Besides the spin relaxation, the SOC results in the coupling between $z$ and $x$ projections of the triplet. Such a coupling is caused by precession of electron's spins relative the direction of the mean spin-orbit field $\mathbf{h}_{\mathbf{k}}$, which is created by electrons diffusing along the $x$-axis and, hence, is parallel to the $y$ axis. At the same time $D_{yz}$ stays decoupled from other components of the triplet. This physics is well known in the case of the spin diffusion in spin-orbit coupled normal metals.\cite{Mishch,Burkov,MalshPRLSHE}

At finite $\xi$ the diffusion equations describe qualitatively new physical effects. First of all, as seen from  Eq.(\ref{Dzz3}), a new channel of the spin relaxation with the relaxation rate $\sim \Gamma \xi^2$ appears. This is the same strong relaxation mechanism which acts on singlet pairing correlations and is represented in  Eq.(\ref{D00}). At $\xi \sim 1$ it primary involves the triplet component parallel to the N\'{e}el vector \cite{Brataas1}. At the same time, there are also long-range diffusion modes originating from triplets  perpendicular to  the N\'{e}el vector. They relax via the D'yakonov-Perel mechanism. At $\xi \sim 1$ the latter may be much  weaker than the relaxation caused by the AF order. Moreover, due to antiferromagnetism the mutual coupling of  triplets  is qualitatively distinct from that in the absence of the N\'{e}el order.  In particular, the term in  the right-hand side of Eq.(\ref{D11}), which is proportional to $\xi$, gives rise to the entanglement of $z$ and $y$ components of the triplet. Therefore, its three components become mutually coupled due to interactions of  electrons with antiferromagnetically ordered spins. Such sort of  entanglement is associated with the interplay of the Rashba coupling and the AF order, as can be seen from Eq.(\ref{D11}), where the mixing term is proportional to $\xi \kappa$. However, the antiferromagnetism generates in the right-hand  side of Eq.(\ref{D11}) the terms $I_{xy}$ and $I_{yx}$ which, as seen from Eq.(\ref{I}), do not depend on the Rashba coupling. It is easy to see from the upper Eq.(\ref{D11}) that such terms give rise to the pairing functions $D_{xy}$ and $D_{yx}$. It is therefore not necessary to engage the Rashba coupling for generating such nondiagonal functions. Moreover, the signs of  these functions are opposite on different sublattices. This results in nontrivial consequences for the Josephson current which will be discussed in the next section.
\section{Josephson current}
\subsection{Contacts}

The interaction of AF with superconducting contacts is represented by equations (\ref{Sigma2}) and (\ref{j}). The integral over $\mathbf{k}^{\prime}$ in Eq.(\ref{Sigma2}) may be expressed in terms of the semiclassical Green function of the superconductor with a thin ferromagnetic film on its surface. For a dirty superconductor this function can be calculated by using the semiclassical Usadel equation. \cite{Bergeret}  By integrating  $F^{s}_{L(R)\mathbf{k}}(\omega_n)$ in Eq.(\ref{Sigma2}) over $\mathbf{k}$ we obtain
\begin{equation}\label{f}
\frac{1}{V_{L(R)}}\sum_{\mathbf{k}^{\prime}}F^{s}_{L(R)\mathbf{k}^{\prime}}(\omega_n)=-i\pi N^s_{\mu L(R)}\rho_zf_{L(R)}(\omega_n)\,,
\end{equation}
where $f_{L}$ and $f_R$ are the anomalous semiclassical Green functions of the left and right contacts and $N^s_{\mu L(R)}$ is the electronic (normal) density of states on the Fermi level. Since the tunneling contacts of ferromagnetic films to AF are weak, one may use the known result \cite{Bergeret,comment} for a ferromagnetic film where one its surface makes contact with the superconductor and the other is open. For simplicity, both contacts will be treated equivalent, except that magnetization directions in ferromagnetic films and phases of  superconducting order parameters can be different. In this case $f_{L(R)}$ has the form
\begin{equation}\label{f2}
f_{L(R)}=(f^s_{L(R)}+\bm{\sigma}\mathbf{f}^t_{L(R)})(\rho^+e^{i\phi_{L(R)}/2}-\rho^-e^{-i\phi_{L(R)}/2})\,,
\end{equation}
where $\phi_{L}=\phi$ and $\phi_{R}=-\phi$ are the phases of the order parameters in the left and right contacts, respectively, and $\rho^{\pm}=\rho_x\pm i\rho_y$. In Eq.(\ref{f2}) the first term corresponds to the singlet part of the Green function, while the second one is associated with the triplet pairing. According to Ref.[\onlinecite{Bergeret}], the latter corresponds to the odd triplet pairing of electrons having the total spin 1 whose projection onto the ferromagnetic magnetization direction is 0. Since this magnetization is homogeneous, only this sort of the triplet pairing takes place in the ferromagnetic film. The corresponding pairing functions are expressed as $2f^s_{L(R)}=f_++f_-$ and $2\mathbf{f}^t_{L(R)}=\mathbf{e}_{L(R)}(f_+-f_-)$, where  $\mathbf{e}_{L(R)}$ is the unit vector parallel to the magnetization of the ferromagnet. In turn, the functions $f_+$ and $f_-$ are given by
\begin{equation}\label{f0t}
f_{\pm}(\omega_n)=-\frac{i\Delta}{\gamma_{\mathrm{fs}}\sqrt{\omega_n^2+\Delta^2}}\frac{1}{\eta_{\pm}\sinh d_f\eta_{\pm}} \,,
\end{equation}
where $\Delta >0$ is the superconducting order parameter, $\gamma_{\mathrm{fs}}$ is a coefficient which is proportional to the resistance of the superconductor - ferromagnet interface, $\eta_{\pm}=\sqrt{2(|\omega_n|\mp i \mathrm{sign}(\omega_n) H_{ex})/D_f}$ and $d_f$ is the thickness of the ferromagnetic film, while $D_f$ and $H_{ex}$ are the diffusion constant
and the energy of the exchange interaction of electrons with localized spins, respectively.
As seen, $f_+$ and $f_-$ decrease exponentially  when $d_f>\sqrt{D_f/H_{ex}}$. As a result, the ferromagnetic film must be sufficiently thin. Usually exchange energy $H_{ex}$ is more than several tens of meV,  that is  much larger than the magnitude of the superconducting energy gap. Therefore, it is reasonable to ignore $|\omega_n|$ in $\eta_{\pm}$. As a result, the frequency dependence of $f_{\pm}(\omega_n)$ in equations Eq.(\ref{f0t}) and Eq.(\ref{f2}) is simplified. So, $f^s_{L(R)}$ and  $\mathbf{f}^t_{L(R)}$ may be represented in the form: $f^s_{L(R)}=-iA_s\Delta/\sqrt{\omega_n^2+\Delta^2}$ and $\mathbf{f}^t_{L(R)}=\mathbf{e}_{L(R)}\mathrm{sign}(\omega_n)A_t\Delta/\sqrt{\omega_n^2+\Delta^2}$, while $A_s$ and $A_t$ are frequency independent dimensionless coefficients given by
\begin{equation}\label{A0t}
A_s= \frac{1}{\gamma_{\mathrm{fs}}}\mathrm{Re}\frac{1}{\eta\sinh d\eta}\,\,,\,A_t=\frac{1}{\gamma_{\mathrm{fs}}}\mathrm{Im}\frac{1}{\eta\sinh d_f\eta}\,,
\end{equation}
where  $\eta\simeq\sqrt{2i  H_{ex}/D_f}$ at $\omega_n \ll H_{ex}$. Note, that the triplet pairing function $\mathbf{f}_{L(R)}$ is proportional to $\mathrm{sign}(\omega_n)$, because it corresponds to the odd-frequency  superconducting pairing. \cite{Bergeret}

Let us consider superconducting contacts which have the same width $w$ in the $x$-direction and the length $l_y$ in the $y$-direction. The shape of the contacts may be taken into account by multiplying the self energies $\Sigma_{L}$ and $\Sigma_{R}$ by the functions $s_{L}(x)=a^{-1}\theta(x+w+d/2)\theta(-x-d/2)$ and $s_{R}(x)=a^{-1}\theta(x-d/2)\theta(w+d/2-x)$, respectively, where the distance between the contacts is $d$. The Fourier components of these functions will be denoted as $s_{q_{xL(R)}}$. If $wq\ll 1$, for such thin contacts, whose width is much smaller than the spin relaxation length and the length of the junction,  $s_{q_{xL(R)}}$ may be written as
\begin{equation}\label{wq}
s_{q_{xL(R)}}=\frac{w}{a}e^{\pm iq_{x}d/2}\,.
\end{equation}

\subsection{Current}

By using equations (\ref{Sigma2}),  (\ref{K}), (\ref{D}) and (\ref{f}-\ref{A0t}) expression Eq.(\ref{j}) for the Josephson current may be written in the form
\begin{eqnarray}\label{j2}
&&j=ie\pi^2N^{s2}_{\mu}N_{\mu}\Delta^2Tl_y\sum_{\omega_n}\int \frac{dq_x }{2\pi}\frac{e^{iq_{x}d}}{\Gamma} \frac{s_{q_{xL}}s_{-q_{xR}}}{\omega_n^2+\Delta^2}\times \nonumber\\
&&\sum_{\lambda\lambda^{\prime}}|t^{\lambda}_{L}|^2|t^{\lambda^{\prime}}_{R}|^2 [
A_s^2(e^{-i\phi}D^{1-1,\lambda\lambda^{\prime},}_{00,q_x}-e^{i\phi}D^{-11,\lambda\lambda^{\prime}}_{00,q_x})-\nonumber \\
&&e_L^ie_R^jA^2_t(e^{-i\phi}D^{1-1,\lambda\lambda^{\prime}}_{ij,q_x}-e^{i\phi}D^{-11,\lambda\lambda^{\prime}}_{ij,q_x})]\,.
\end{eqnarray}
Within the considered semiclassical approximation the singlet current is decoupled from the triplet one. Therefore, in Eq.(\ref{j2}) the correlation functions  $D_{0j}$ and $D_{i0}$  are absent. It is seen from Eq.(\ref{D00}) that the singlet pairing function $D_{00}$ does not depend on sublattice and Nambu variables. Hence, it follows from Eq.(\ref{j2}) that the corresponding current is proportional to $\sin\phi$. At the  fixed frequency the integration of   $D_{00}$ over $q_x$ in Eq.(\ref{j2}) results in an exponential decreasing of the current with the junction length. This is distinct from ferromagnetic junctions, \cite{Buzdin, Bergeret} where the critical current oscillates with the junction length and exchange field. Such a simple exponential behavior takes place only if $\xi\ll 1$. Otherwise, the diffusion approximation which was employed to derive Eq.(\ref{D00}) is not valid.

At the same time, the current, which is associated with triplet pairing functions, is not so trivial, as its singlet counterpart. Let us first consider the currents  $j_{xx}$ and $j_{yy}$, which are produced by  $D_{xx}$  and $D_{yy}$ functions in Eq.(\ref{j2}), respectively. When $\xi \sim 1$,  Eq.(\ref{I}) and the upper equation in Eq.(\ref{D11}) give at $i=j$
\begin{equation}\label{Dyy}
D^{11}_{xx(yy)}=-\Gamma(1+\xi^2)(2|\omega_n|+c_1Dq_x^2+4c_1\kappa^2D)^{-1}
\end{equation}
and
\begin{equation}\label{Dyy12}
D^{12}_{xx(yy)}=-\Gamma(1-\xi^2)(2|\omega_n|+c_1Dq_x^2+4c_1\kappa^2D)^{-1}\,,
\end{equation}
where only sublattice  superscripts are shown, while the Nambu ones are ignored, because the diagonal functions $D^{\lambda\lambda^{\prime}}_{ii}$ ($i\in (x,y)$) do not depend on Nambu variables. We have also $D^{22}_{ii}=D^{11}_{ii}$ and $D^{21}_{ii}=D^{12}_{ii}$.  Therefore,  Eq.(\ref{j2}) results in the $\sin\phi$ phase dependence. Similar to the singlet supercurrent, the triplet currents, whose spins are oriented along $x$ and $y$ directions also decrease with the junction length. But, unlike the singlet current, at $\Delta \ll D\kappa^2$ this decreasing  stems mostly  from the D'yakonov-Perel' spin relaxation which is represented by the third term in the left-hand side of Eq.(\ref{D11}). However, since the typical characteristic spin-orbit length $\kappa^{-1}$ may be larger, or of the same order of magnitude as the junction length, the triplet supercurrents with spins oriented perpendicular to the N\'{e}el vector can flow through relatively long junctions, even at $\xi \sim 1$.

In contrast,  the triplet current whose spin  is parallel to the $z$ axis decreases slowly  with $d$  only in the case when $\xi$ is sufficiently small. In the leading approximation at $\xi \ll 1$ the corresponding function $D_{zz}$ in Eq.(\ref{j2}) can be obtained from Eq.(\ref{Dzz3}). It can be seen from this equation that both, the D'yakonov-Perel' spin relaxation and  the elastic scattering from potential disorder in an AF ordered metal, contribute to the spin relaxation. These mechanisms give rise to a decreasing of the current with the junction length. At the fixed frequency the corresponding exponent is given by $d(\Gamma D)^{-1/2}\sqrt{2\Gamma|\omega|+2\langle h^2_{\mathbf{k}}\rangle+4\Gamma^2\xi^2}$. It is seen that when $\langle h^2_{\mathbf{k}}\rangle^{1/2}\lesssim \Gamma\xi$  and $\Delta \lesssim \Gamma\xi^2$ the current decreases fast at $2\xi d\sqrt{\Gamma/D}\gtrsim 1$, similar to the discussed above singlet supercurrent.

Let us now consider the situation when magnetizations of superconducting contacts are not collinear. If magnetizations of the left and right contacts are parallel to the $x$ and $y$ axes, respectively, the Josephson current in Eq.(\ref{j2}) is determined by  $D^{\rho\overline{\rho},\lambda\lambda^{\prime}}_{xy}$. The latter can be obtained from Eq.(\ref{D11}) and Eq.(\ref{I}) and has the form
\begin{equation}\label{Dxy}
D^{\rho\overline{\rho},11(22)}_{xy}=-D^{\rho\overline{\rho},11(22)}_{yx}=\pm\frac{2i\rho\Gamma \xi}{2|\omega_n|+c_1Dq_x^2+4c_1\kappa^2D}
\end{equation}
and
\begin{equation}\label{Dxy12}
D^{\rho\overline{\rho},12}_{xy}=-D^{\rho\overline{\rho},12}_{yx}=0\,.
\end{equation}
It follows from Eq.(\ref{j2}) that the Josephson current, which is associated with these functions, is proportional to $\cos\phi$ because they change their signs at $\rho \rightarrow -\rho$. Such a current does not reverses its direction at $\phi \rightarrow -\phi$, as it is required by the time reversal symmetry. This can be explained by a sublattice selectivity of considered correlation functions. Each sublattice carries a ferromagnetically ordered localized spins which flip their orientation at time reversal. Besides $D_{xy}$, other correlation functions can have such a sublattice selectivity and may give rise to $\cos\phi$ phase dependence of the current. Let us consider the functions $D_{xz}$ and  $D_{yz}$. They can be obtained from Eq.(\ref{D11}). At $\xi \sim 1$, by substituting Eq.(\ref{Dzz2}) in Eq.(\ref{D11}) one can see that  $D_{xz}$ is symmetric with respect to $\xi$, while $D_{yz}$ changes its sign with $\xi$. In the latter case Eq.(\ref{D11}) and Eq.(\ref{Dzz2}) give
\begin{equation}\label{Dyz}
D^{\rho\overline{\rho},11}_{yz}=D^{\rho\overline{\rho},12}_{yz}=-\frac{c_2\rho\xi^{-1} \kappa Dq_x(1+\xi^{2})}{2|\omega_n|+c_1Dq_x^2+4c_1\kappa^2D}\,.
\end{equation}
At the same time $D^{\rho\overline{\rho},11}_{xz}$ is given by
\begin{equation}\label{Dxz}
D^{\rho\overline{\rho},11}_{xz}=D^{\rho\overline{\rho},12}_{xz}=-i\rho\xi D^{\rho\overline{\rho},11}_{yz}
\end{equation}
By reversing the sign of $\xi$ in equations Eq.(\ref{Dyz}) and Eq.(\ref{Dxz}) one can obtain the functions $D^{22}_{iz}$ and $D^{21}_{iz}$ ($i=x,y$). It is seen that $D^{\rho\overline{\rho},\lambda\lambda^{\prime}}_{xz}, D^{\rho\overline{\rho},\lambda\lambda^{\prime}}_{yz}$, and $D^{\rho\overline{\rho},\lambda\lambda^{\prime}}_{xy}$ decrease relatively slowly with $d$ due to the D'yakonov-Perel' mechanism, while the strong destructive effect, which is caused by antiferromagnetism, is turned off. This behavior is similar to the current associated  with the correlators $D^{\lambda\lambda^{\prime}}_{xx(yy)}$, which are given by Eq.(\ref{Dyy}) and Eq.(\ref{Dyy12}). Note, that unlike the functions $D_{xz}$ and  $D_{yz}$ turning to zero with vanishing  Rashba SOC, the diffusion propagator $D_{xy}$ stays finite. On this reason it gives a much stronger contribution to the  current. At $\xi \sim 1$, by comparing Eq.(\ref{Dxy}) and Eq.(\ref{Dyz}), it is easy to see that the ratio of these functions is of the order of $D\kappa^2/\Gamma \sim l_i^2\kappa^2 \ll 1$, where $l_i$ is the electron's mean free path.

Because of  sublattice selectivity of triplet's  diffusion, it is important to consider the sublattice dependence of tunneling constants. A simplest case is when these parameters do not depend on the sublattice index, i.e. we have $|t^1_{L(R)}|^2=|t^2_{L(R)}|^2\equiv |t_{L(R)}|^2$. Then, it follows from Eq.(\ref{j2}) that the current is expressed via the sum $D^{11}+D^{12}+D^{22}+D^{21}$. In this case, only those terms which are symmetric with respect to $\xi$ contribute to the sum, because $D^{11}(\xi)+D^{12}(\xi)=D^{22}(-\xi)+D^{21}(-\xi)$.

In general, one can transform Eq.(\ref{j2}) by taking into account the Hermitian character of  the matrix $D$ and by using relations between $D^{\rho\bar{\rho}}$ and $D^{\bar{\rho}\rho}$ which are obtained  in Abstract A. So, the expression in the third line of Eq.(\ref{j2}) can be written as
\begin{eqnarray}\label{rhobarrho}
&&e^{-i\phi}D^{\rho\overline{\rho},\lambda\lambda^{\prime}}_{ij,q_x}-e^{i\phi}D^{\rho\overline{\rho},\lambda\lambda^{\prime}}_{ij,q_x}=\nonumber \\
&&e^{-i\phi}D^{\rho\overline{\rho},\lambda\lambda^{\prime}}_{ij,q_x}-e^{i\phi}D^{\rho\overline{\rho},\lambda\lambda^{\prime}*}_{ij,-q_x}\,.
\end{eqnarray}
From this equation it is immediately seen that, if $D^{\rho\overline{\rho},\lambda\lambda^{\prime}}_{ij,q_x}$ is an even and real function of $q_x$, one arrives at the usual $\sin\phi$ dependence of the current. This result does not depend on sublattice and spin variables of $D^{\rho\overline{\rho},\lambda\lambda^{\prime}}_{ij,q_x}$. At the same time, if $D^{\rho\overline{\rho},\lambda\lambda^{\prime}}_{ij,q_x}$ is an even and imaginary function of $q_x$, its contribution to the current is proportional to $\cos\phi$. Such a phase dependence takes place for the  function $D^{\rho\overline{\rho},\lambda\lambda^{\prime}}_{xy}$ which is given by Eq.(\ref{Dxy}). It enters in Eq.(\ref{j2}) as the combination $|t^{1}_{L}|^2|t^{1}_{R}|^2D^{\rho\overline{\rho},\lambda\lambda^{\prime}}_{xy} +|t^{2}_{L}|^2|t^{2}_{R}|^2 D^{\rho\overline{\rho},\lambda\lambda^{\prime}}_{xy}$. Since  $D^{\rho\overline{\rho},11}_{xy}=-D^{\rho\overline{\rho},22}_{xy}$, the latter sum becomes $(|t^{1}_{L}|^2|t^{1}_{R}|^2 - |t^{2}_{L}|^2|t^{2}_{R}|^2)D^{\rho\overline{\rho},11}_{xy}$. Therefore, at least one of two terminals must make a  more strong contact with a selected sublattice to reach the finite  current $j_{xy}$. The correlation function $D^{\rho\overline{\rho},\lambda\lambda^{\prime}}_{yz}$, as follows from Eq.(\ref{Dyz}), is the odd with respect to $q_x$ and real function. As a result, it produces the nonreciprocal current $j_{yz}$, whose sublattice dependence is similar to  $j_{xy}$.  In contrast to $D^{\rho\overline{\rho},\lambda\lambda^{\prime}}_{yz}$, the function $D^{\rho\overline{\rho},\lambda\lambda^{\prime}}_{xz}$, which is given by Eq.(\ref{Dxz}), is the odd with respect to $q_x$ imaginary function. Therefore, it results in the current having the usual $\sin\phi$ phase dependence. This current does not turn to zero when transmission coefficients are independent on sublattice variables.
\begin{figure}[tp]
\includegraphics[width=6cm]{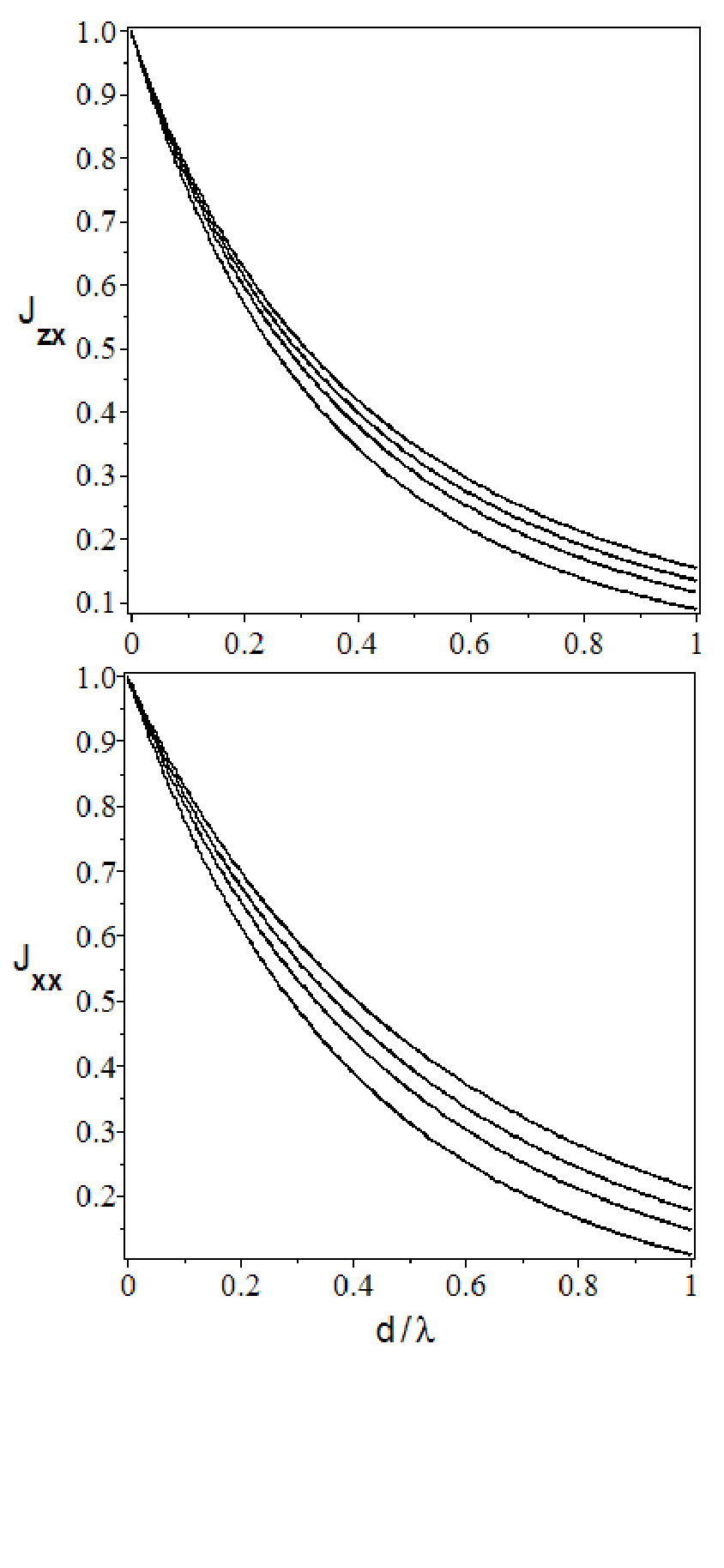}
\caption{The dependence of the critical current on the distance between contacts. The distance is measured in units of $\lambda=\sqrt{D/\Delta}$. The current is normalized by its magnitude at $d=0$. Top: the magnetization of the left(right) contact is  parallel to the z-axis, while that of the right(left) contact is  parallel to the x-axis. Bottom: magnetizations of both contacts are parallel to the x-axis The curves are calculated at $\pi k_BT/\Delta=0.3$ and various strengths of the SOC coupling, which are (from top to bottom): $\kappa\lambda$=0.1, 0.5, 1.0, 2.0} \label{fig3}
\end{figure}

Josephson's current dependence on the distance between contacts is controlled by the following parameters, which have the dimension of length: $\sqrt{\Delta/D}$, $\kappa^{-1}$,   and $l_i/\xi$. The latter parameter is the smallest one, as long as $\xi \sim 1$. Therefore, at $d\gg l_i/\xi$ the singlet current and the current of triplets, whose spins are parallel to the N\'{e}el order, are strongly reduced, as it was discussed above. The dependencies on $d$ of the  triplet currents $j_{xx(yy)}$ and $j_{xy}$ are identical, as well as are identical the  dependencies of $j_{yz}$ and $j_{xz}$. These results are shown at Fig.3. The calculated curves do not show the absolute current magnitude, but only its dependence on the junction length. It is seen that despite different dependencies of $j_{xx}$ and $j_{zx}$ on the wave vector, the curves look rather similar. That is, because in the considered parameter range the D'ykonov-Perel' spin relaxation dominates, so, in both cases currents demonstrate the dependence on $d$ which is close to the exponential one.

\section{Discussion}

In summary, the Josephson current through 2D AF metal has been calculated at various magnetizations of  s-wave superconducting contacts. At $\xi \sim 1$ the current which is created by Cooper pairs with zero spin is weak and can be ignored, if the distance between contacts is considerably larger than the mean free path of electrons. The same situation takes place for the current of triplets whose spins are parallel to the N\'{e}el order. At the same time, triplet Cooper pairs, which are polarized perpendicular to the N\'{e}el order, may contribute to the current, if the D'yakonov-Perel' spin relaxation rate $\langle h^2_{\mathbf{k}}\rangle/\Gamma\lesssim E_{Th}$, where $E_{Th}=D/d^2$ is the Thouless energy. Depending on magnetizations of superconducting contacts,  the Josephson currents $j_{ij}$ and $j_{iz}$, where $(i,j)\in (x,y)$, have been calculated. In these currents the first and second subscripts are related to the left and right contacts, respectively and denote their magnetization directions. Within the considered model the diagonal currents $j_{xx}$ and $j_{yy}$ are identical and demonstrate the conventional $\sin\phi$ superconducting phase dependence. The same dependence takes place for $j_{xz}$. This current is associated with precession of the triplet spin within the $xz$ plane, when a correlated pair of electrons diffuses in the $x$-direction between terminals. This precession is caused by the Rashba SOC. The same evolution of the spin density is well known for the electron diffusion  in normal systems. In contrast, the current $j_{yz}$ behaves in a quite different way. In this case the spin of a triplet pair evolves in the $yz$ plane due to a combined effect of the SOC and the N\'{e}el order. This rotation is sublattice selective. So,  the spin precession occurs  in opposite directions in different sublattices. Such a  current is nonreciprocal, because it is proportional to $\cos\phi$. However, it is not zero only if superconducting terminals contact selectively to AF sublattices. Otherwise, the current is compensated due to opposite directions of the spin precession in different sublattices. One more nonreciprocal current occurs at the $xy$ configuration of terminal's magnetizations. This current is much stronger than $j_{yz}$, because it does not vanish when the Rashba coupling turns to zero. Similar to the $xz$ current, it can be observed only when the contacts selectively interact with AF sublattices. If such a selectivity might be realized in practice, the considered S-AFM-S system could be employed as a $\varphi_0$ junction with $\varphi_0=\pm\pi/2$. Moreover, since the singlet current is strongly suppressed in disordered AF, only triplet currents are represented in Eq.(\ref{j2}). Therefore, the two configurations of contacts, $xy$ and $yz$, guarantee the unidirectional current independently on the sign of the phase difference between superconducting leads.

Within the considered model, sublattices are equivalent and tunneling parameters are independent on the sublattice variable. In this case only the $xx$, $yy$ and $xz$ configurations contribute to the Josephson current. There are several ways to reach the sublattice selectivity of contacts. One can use a compensated ferrimagnet instead of AFM . Since there different atoms occupy  sublattices, their interactions with superconducting contacts may be different. Another way is to employ bilayer systems, where two ferromagnetic layers interact antiferromagnetically and form an AF system. It is also possible to consider systems where AF and orbital orders are correlated, similar to altermagnets.  In all these cases an appropriate model should be used. Such a study is outside the scope of the present work.

In this work we considered the thermally equilibrium Josephson current. At the same time, in terms of the superconducting spintronics, it is interesting to study various processes which involve spin transport. For instance, spin wave modes in AFM have different weights on two sublattices. Therefore, they might affect the spin selective transport of Cooper pairs in Josephson junctions and nontrivially manifest themselves in the current. Also, it is interesting to study effects of dynamical variations of magnetization in ferromagnetic contacts on the Josephson current.

\appendix

\section{Calculation of  the two-particle correlation function}

The function $\Pi^{\rho\bar{\rho},\lambda\lambda^{\prime}}_{ij,\mathbf{q}}(\omega_n)$ is given by Eq.(\ref{Piij}). As was explained above, since a weak spin-charge coupling is ignored,  the singlet two-particle correlator $\Pi^{\rho\bar{\rho},\lambda\lambda^{\prime}}_{00,\mathbf{q}}(\omega_n)$ is decoupled from the triplet one, where $i$ and $j$ take on the values $x,y$ and $z$. By substituting Eq.(\ref{G}) in Eq.(\ref{Piij}) and expanding Eq.(\ref{Piij}) up to  $q^2$ terms, the singlet correlator can be expressed as
\begin{equation}\label{Pi00}
\Pi^{\rho\bar{\rho},\lambda\lambda^{\prime}}_{00,\mathbf{q}}=-\frac{\Gamma}{2W}(1-\xi^2)(1-\frac{\langle v^2\rangle q^2}{8W^2})\,,
\end{equation}
where $\xi=J/\mu$, $W=|\omega_n|+\tilde{\Gamma}$, $\langle v^2\rangle= \sum_{\mathbf{k}}\delta(E_{\mathbf{k}}-|\mu|)v^2/\sum_{\mathbf{k}}\delta(E_{\mathbf{k}}-|\mu|)$, and the electron velocity is given by $\mathbf{v}=\nabla E_{\mathbf{k}}$. Eq.(\ref{Pi00}) is valid for all combinations of the sublattice variables $\lambda,\lambda^{\prime}$ and the signs of $\rho$. It is easy to check that mixed terms of the form $\Pi_{0i}$, or $\Pi_{i0}$, where $i\neq 0$, turn to zero. On this reason the singlet and triplet components of two-particle correlations are decoupled. In contrast, triplet superconducting correlations are coupled to each other. Therefore, diagonal, as well as nondiagonal elements of the matrix $\Pi_{ij,\mathbf{q}}(\omega_n)$ take place. From Eq.(\ref{Piij})  the diagonal terms are obtained in the form
\begin{eqnarray}\label{Pixx}
&&\Pi^{\rho\bar{\rho},11}_{xx,\mathbf{q}}=
\Pi^{\rho\bar{\rho},11}_{yy,\mathbf{q}}=\Pi^{\rho\bar{\rho},22}_{xx,\mathbf{q}}=\Pi^{\rho\bar{\rho},22}_{yy,\mathbf{q}}=\nonumber \\&&-\frac{\Gamma}{2W}(1+\xi^2)\left(1-\frac{\langle v^2\rangle q^2}{8W^2}-\frac{h^2_{\mathbf{k}}}{2W^2}\right)\,
\end{eqnarray}
and
\begin{eqnarray}\label{Pixx12}
&&\Pi^{\rho\bar{\rho},12}_{xx,\mathbf{q}}=
\Pi^{\rho\bar{\rho},12}_{yy,\mathbf{q}}=\Pi^{\rho\bar{\rho},21}_{xx,\mathbf{q}}=\Pi^{\rho\bar{\rho},21}_{yy,\mathbf{q}}=\nonumber \\&&-\frac{\Gamma}{2W}(1-\xi^2)\left(1-\frac{\langle v^2\rangle q^2}{8W^2}-\frac{h^2_{\mathbf{k}}}{2W^2}\right)\,.
\end{eqnarray}
These equations are valid for any $\rho=\pm 1$. In turn, the function $\Pi_{zz}$ is given by
\begin{equation}\label{Pizz}
\Pi^{\rho\bar{\rho},\lambda\lambda^{\prime}}_{zz,\mathbf{q}}=-\frac{\Gamma}{2W}(1-\xi^2)\left(1-\frac{\langle v^2\rangle q^2}{8W^2}-\frac{h^2_{\mathbf{k}}}{W^2}\right)\,.
\end{equation}
The latter equation is valid for all combinations of sublattice variables $\lambda$ and $\lambda^{\prime}$, as well as for all signs of $\rho$. For nondiagonal matrix elements we obtain
\begin{eqnarray}\label{Pixy}
&&\Pi^{\rho\bar{\rho},11}_{xy,\mathbf{q}}=
-\Pi^{\rho\bar{\rho},11}_{yx,\mathbf{q}}=-\Pi^{\rho\bar{\rho},22}_{xy,\mathbf{q}}=\Pi^{\rho\bar{\rho},22}_{yx,\mathbf{q}}=\nonumber \\&&i\xi\rho\frac{\Gamma}{W}\left(1-\frac{\langle v^2\rangle q^2}{4W^2}-\frac{h^2_{\mathbf{k}}}{2W^2}\right)
\end{eqnarray}
and $\Pi^{\rho\bar{\rho},12}_{xy,\mathbf{q}}=\Pi^{\rho\bar{\rho},12}_{yx,\mathbf{q}}=0$. The same is valid for $\Pi^{21}$ matrix elements. It is important that the functions in Eq.(\ref{Pixy}) change their signs at $\rho\rightarrow -\rho$, in contrast to diagonal $xx$,$yy$ and $zz$ terms. Other nondiagonal functions involve mixed $xz$ and $yz$ matrix elements. For brevity, they are explicitly presented at $\rho=1$ and Nambu superscripts are suppressed. From Eq.(\ref{Piij})  these functions  are obtained in the form
\begin{eqnarray}\label{Pixz}
&&\Pi^{11}_{xz,\mathbf{q}}=(\Pi^{11}_{zx,\mathbf{q}})^*=\Pi^{12}_{xz,\mathbf{q}}=-\Pi^{12}_{zx,\mathbf{q}}=\nonumber\\
&&-\frac{\Gamma}{2W^3}\sqrt{1-\xi^2}(\xi\langle v_yh^x_{\mathbf{k}}\rangle q_y-i\langle v_xh^y_{\mathbf{k}}\rangle q_x)
\end{eqnarray}
and
\begin{eqnarray}\label{Piyz}
&&\Pi^{11}_{yz,\mathbf{q}}=(\Pi^{11}_{zy,\mathbf{q}})^*=\Pi^{12}_{yz,\mathbf{q}}=-\Pi^{12}_{zy,\mathbf{q}}=\nonumber\\
&&-\frac{\Gamma}{2W^3}\sqrt{1-\xi^2}(\xi\langle v_xh^y_{\mathbf{k}}\rangle q_x+i\langle v_yh^x_{\mathbf{k}}\rangle q_y)\,.
\end{eqnarray}
Expressions for $\Pi^{22}$ and $\Pi^{21}$ can be obtained from Eqs.(\ref{Pixz}) and (\ref{Piyz}) by the substitution $1 \rightleftarrows 2$ and $ J \rightarrow -J$. The above equations for nondiagonal matrix elements are written at Nambu index $\rho=1$. At $\rho=-1$ the corresponding functions can be obtained by changing in Eq.(\ref{Piij}) $\rho \rightarrow -\rho$, $i \rightleftarrows j$ and $\mathbf{q}\rightarrow - \mathbf{q}$ and by permutations under the trace operation. So, we arrive at $\Pi^{\bar{\rho}\rho,\lambda\lambda^{\prime}}_{ij,\mathbf{q}}=\Pi^{\rho\bar{\rho},\lambda^{\prime}\lambda}_{ji,-\mathbf{q}}$. One can also directly derive from Eq.(\ref{Piij}) the helpful Hermitian relation $\Pi^{\rho\bar{\rho},\lambda\lambda^{\prime}}_{ij,\mathbf{q}}=(\Pi^{\rho\bar{\rho},\lambda^{\prime}\lambda}_{ji,\mathbf{q}})^{*}$.

\section{Derivation of the diffusion equations}

A most general form of equations for the function $D_{\mathbf{q}}(\omega_n)$ is given by Eq.(\ref{K2}).   These equations can be written as
\begin{eqnarray}\label{K3}
&&D^{11}_{ij}=\Pi^{11}_{ij}-\Pi^{11}_{il}D^{11}_{lj}-\Pi^{12}_{il}D^{21}_{lj}\,,\nonumber \\
&&D^{21}_{ij}=\Pi^{21}_{ij}-\Pi^{22}_{il}D^{21}_{lj}-\Pi^{21}_{il}D^{11}_{lj}\,,\nonumber\\
&&D^{22}_{ij}=\Pi^{22}_{ij}-\Pi^{22}_{il}D^{22}_{lj}-\Pi^{21}_{il}D^{12}_{lj}\,,\nonumber \\
&&D^{12}_{ij}=\Pi^{12}_{ij}-\Pi^{11}_{il}D^{12}_{lj}-\Pi^{12}_{il}D^{22}_{lj}\,,
\end{eqnarray}
where the frequency and wave vector dependencies of the functions $\Pi$ are not shown. Nambu superscripts are also suppressed in this section. As follows from Eq.(\ref{Pixz}) and Eq.(\ref{Piyz}) the functions $\Pi_{iz}$ and $\Pi_{zi}$, where $i\in (x,y)$ are small. They are proportional to $q$ and $h_{k}$ and, hence, carry the small parameter $Dq\kappa/\Gamma \ll 1$. Therefore, $(x,y)$ and $(z)$ subspaces in Eq.(\ref{K3}) are weakly coupled to each other. In the leading approximation it is possible to calculate separately $D_{zz}$ and $D_{ij}$, where $(i,j)\in(x,y)$.  In this approximation $D_{zz}$ was calculated in Sec.IIC. In turn, the equation for $D_{ij}$ is truncated Eq.(\ref{K3}), which is restricted to the $(x,y)$ subspace. From these equations one can obtain expressions for sublattice matrix elements. For a pair $D^{11}$ and $D^{12}$ we get
\begin{eqnarray}\label{K4}
&&MD^{11}=-(1+\Pi^{22})+M\,,\nonumber \\
&&MD^{12}=(1+\Pi^{22})\Pi^{12}-\nonumber \\
&&(1+\Pi^{22})\Pi^{12}(1+\Pi^{22})^{-1}\Pi^{22}\,,
\end{eqnarray}
where
\begin{eqnarray}\label{M}
&&M=(1+\Pi^{22})(1+\Pi^{11})-\nonumber\\
&&(1+\Pi^{22})\Pi^{12}(1+\Pi^{22})^{-1}\Pi^{21}\,.
\end{eqnarray}
By substituting in Eq.(\ref{M}) expressions for the matrix elements $\Pi_{ij}$ from equations Eq.(\ref{Pixx}),Eq.(\ref{Pixx12}) and Eq.(\ref{Pixy}) we arrive at
\begin{eqnarray}\label{M2}
&&M_{xx(yy)}=\frac{c_1}{2\Gamma}\left(2|\omega_n|+\frac{c_1\langle v^2\rangle q^2}{4\Gamma}+\frac{c_1\langle h^2_{\mathbf{k}}\rangle}{\Gamma}\right) \,, \nonumber\\
&&M_{xy}=M_{yx}=0\,,
\end{eqnarray}
where $c_1=1/(1+\xi^2)$. By taking into account only the leading terms in Eq.(\ref{K4}) one may neglect in its  right-hand side all small terms which are proportional to $q, h$ and $\omega_n$. Let us denote the  source terms for $D^{11}$ and $D^{12}$, which are represented by the right-hand sides of  Eq.(\ref{K4}), as $I^{11}$ and $I^{12}$, respectively. Then, by using equations in Abstract A we find
\begin{eqnarray}\label{I}
&&I^{12}_{xy}=I^{12}_{yx}=0\,\,\,,\,\,I^{11}_{xy}=-I^{11}_{yx}=i\frac{\xi}{1+\xi^2}\,,\nonumber\\
&&I^{12}_{xx(yy)}=-\frac{1}{2}\frac{1-\xi^2}{1+\xi^2}\,\,\,,\,\,\,I^{11}_{xx(yy)}=-\frac{1}{2}\,.
\end{eqnarray}
The parameters $I^{22}$ and $I^{21}$ can be obtained from Eq.(\ref{I})\ via the substitution $1\rightleftarrows 2$ and $\xi \rightarrow -\xi$.

In order to calculate the correlation functions  $D_{iz}$, where $i\in(x,y)$, let us set in Eq.(\ref{K3}) $j=z$. Then, these equations may be written as
\begin{equation}\label{Kiz}
D_{iz}=\Pi_{iz}(1-D_{zz})-\Pi_{il}D_{lz}\,,
\end{equation}
where sublattice superscripts are suppressed. The function $D_{zz}$ in the right-hand side is calculated within the approximation of decoupled
$(z)$ and $(x,y)$ subspaces. It is given by Eq.(\ref{Dzz2}) and Eq.(\ref{Dzz3}). Since Eq.(\ref{Kiz}) is restricted to the $(x,y)$ subspace, we arrive at Eq.(\ref{K4}) with the modified right-hand side, where $\Pi_{iz}$ is substituted for $\tilde{\Pi}_{iz}=\Pi_{iz}(1-D_{zz})$. The sublattice matrix elements of the latter are $\tilde{\Pi}^{11}_{iz}=\Pi^{11}_{iz}(1-D^{11}_{zz})-\Pi^{12}_{iz}D^{21}_{zz}$ and $\tilde{\Pi}^{12}_{iz}=\Pi^{12}_{iz}(1-D^{22}_{zz})-\Pi^{11}_{iz}D^{12}_{zz}$. Further, let us take into account that $D_{zz}$ does not depend on sublattice variables. Also, according to  Eq.(\ref{Pixz}) and Eq.(\ref{Piyz}), we have $\Pi^{12}_{iz}=\Pi^{11}_{iz}$. Therefore, $\tilde{\Pi}^{11(12)}_{iz}=\Pi^{11}_{iz}(1-2D_{zz})$. By taking into account that, according to Eq.(\ref{M2}), $M_{xy}=M_{yx}=0$,  the modified Eq.(\ref{K4}) for $D_{iz}$ takes the form
\begin{eqnarray}\label{MDiz}
&&M_{ii}D^{11}_{iz}=(\delta_{ij}+\Pi^{22}_{ij})\tilde{\Pi}_{jz}^{11}-\Pi^{12}_{ii}\tilde{\Pi}^{21}_{iz}\,,\nonumber\\
&&M_{ii}D^{12}_{iz}=(\delta_{ij}+\Pi^{22}_{ij})\tilde{\Pi}_{jz}^{12}-\Pi^{12}_{ii}\tilde{\Pi}^{22}_{iz}
\end{eqnarray}
It is easy to see that equations for $D^{11}_{iz}$ and  $D^{12}_{iz}$ are identical. It follows from Eq.(\ref{Pixz}) and Eq.(\ref{Piyz}), in particular, from the relations $\Pi^{12}_{iz}=\Pi^{11}_{iz}$,  $\Pi^{21}_{iz}=\Pi^{22}_{iz}$, and  $\Pi^{12}_{ij}\sim \delta_{ij}$ at $(i,j)\in(x,y)$.  Therefore, $D^{11}_{iz}=D^{12}_{iz}$. From Eq.(\ref{Pixx}), Eq.(\ref{Pixx12}) and Eq.(\ref{Pixy}), where $W\approx \Gamma(1+\xi^2)$, the right-hand side of Eq.(\ref{MDiz}) is obtained in the form
\begin{equation}\label{Iiz}
I^{11}_{iz}=I^{12}_{iz}=\Pi^{11}_{iz}(1-2D_{zz})\,.
\end{equation}
At $\xi \sim 1$ Eq.(\ref{Dzz2}) gives $1-2D_{zz}=(1+\xi^2)/2\xi^2$. Equations for $D^{22}_{\mathbf{q}}(\omega_n)$ and $D^{21}_{\mathbf{q}}(\omega_n)$ can be obtained by the substitutions $1\rightleftarrows 2$ and $J \rightarrow -J$ ($\xi \rightarrow -\xi$) in equations Eq.(\ref{K4}) and Eq.(\ref{M}).

\begin{thebibliography}{99}

\bibitem{Linder1}
J. Linder and A. V. Balatsky, Odd-frequency superconductivity, Rev. Mod. Phys. \textbf{91}, 045005 (2019).
\bibitem{Buzdin}
A. I. Buzdin, Proximity effects in superconductor-ferromagnet heterostructures, Rev. Mod. Phys, \textbf{77}, 935 (2005)
\bibitem{Bergeret}
F. S. Bergeret, A. F. Volkov, and K. B. Efetov, Odd triplet superconductivity and related phenomena in superconductor-ferromagnet structures, Rev. Mod. Phys. \textbf{77}, 1321 (2005).
\bibitem{Robinson}
J. Linder and J. W. A. Robinson, Superconducting spintronics, Nat. Phys. \textbf{11}, 307 (2015)
\bibitem{Qi}
X. L. Qi and S. C. Zhang, Topological insulators and superconductors, Rev. Mod. Phys. \textbf{83}, 1057 (2011)
\bibitem{Baltz}
V. Baltz, A. Manchon, M. Tsoi, T. Moriyama, T. Ono, Y. Tserkovnyak, Antiferromagnetic spintronics, Rev. Mod. Phys. \textbf{90}, 015005 (2018).
\bibitem{Gomonay}
O. Gomonay, V. Baltz, A. Brataas, Y. Tserkovnyak, Antiferromagnetic spin textures and dynamics, Nat. Phys.  \textbf{14} 213 (2018)
\bibitem{Yan}
H. Yan, Z. Feng, P. Qin, X. Zhou, H. Guo, X. Wang, H. Chen, X. Zhang, H. Wu, C. Jiang, Z. Liu, Electric-Field-Controlled Antiferromagnetic Spintronic Devices, Adv. Materials \textbf{32}, 1905603 (2020)
\bibitem{Wadley}
P. Wadley, B. Howells, J. \v{Z}elezn\'{y}, C. Andrews, V. Hills, R. P. Campion, V. Novak, K. Olejnik, F. Maccherozzi, S. S. Dhesi, S. Y. Martin, T. Wagner, J. Wunderlich, F. Freimuth, Y. Mokrousov, J. Kunes, J. S. Chauhan, M. J. Grzybowski, A. W. Rushforth, K. W. Edmonds, B. L. Gallagher, T. Jungwirth, Electrical switching of an antiferromagnet, Science, \textbf{351}, 587 (2016).
\bibitem{Krivoruchko}
V. N. Krivoruchko, Upper critical fields of the superconducting state of a superconductor-antiferromagnetic metal superlattice, JETP \textbf{82}, 347 (1996),  [Zh. Eksp. Teor. Fiz. 109, 649 (1996)]
\bibitem{Linder2021}
L. G. Johnsen, S. H. Jacobsen, and J. Linder, Magnetic control of superconducting heterostructures using compensated antiferromagnets, Phys. Rev. B \textbf{103}, L060505 (2021).
\bibitem{Wu}
B. L. Wu, Y. M. Yang, Z. B. Guo, Y. H. Wu, and J. J. Qiu, Suppression of superconductivity in Nb by IrMn in IrMn/Nb bilayers, Appl. Phys. Lett. \textbf{103}, 152602 (2013)
\bibitem{Hubener}
M. Hubener, D. Tikhonov, I. A. Garifullin, K. Westerholt, and H. Zabel, The antiferromagnet/superconductor proximity effect in Cr/V/Cr trilayers, J. Phys.: Condens. Matter \textbf{14}, 8687 (2002).
\bibitem{Bobkov2}
G. A. Bobkov, I. V. Bobkova, A. M. Bobkov, and A. Kamra, Neel proximity effect at antiferromagnet/superconductor interfaces, Phys. Rev. B \textbf{106}, 144512 (2022).
\bibitem{Brataas1}
E. H. Fyhn, A. Brataas, A. Qaiumzadeh, and J. Linder, Superconducting proximity effect and long-ranged triplets in dirty metallic antiferromagnets,  Phys. Rev. Lett. \textbf{131}, 076001 (2023)
\bibitem{Bobkov1}
G.A. Bobkov, I. V. Bobkova, and A. M. Bobkov, Proximity effect in superconductor/antiferromagnet hybrids: N\'{e}el triplets and impurity suppression of superconductivity, Phys. Rev. B \textbf{108}, 054510 (2023)
\bibitem{Bobkova}
I.V. Bobkova, P. J. Hirschfeld, and Yu. S. Barash, Spin-Dependent Quasiparticle Reflection and Bound States at Interfaces with Itinerant Antiferromagnets, Phys. Rev. Lett \textbf{94}, 037005 (2005)
\bibitem{Barash2005}
B. M. Andersen, I. V. Bobkova, P. J. Hirschfeld, and Y. S. Barash, Bound states at the interface between antiferromagnets and superconductors, Phys. Rev. B \textbf{72}, 184510 (2005).
\bibitem{Bell}
C. Bell, E. J. Tarte, G. Burnell, C. W. Leung, D.-J. Kang, M. G. Blamire, Proximity and Josephson effects in superconductor - antiferromagnetic Nb/-Fe$_{50}$Mn$_{50}$ heterostructures, Phys. Rev. B \textbf{68}, 144517 (2003)
\bibitem{Weides}
 M. Weides, M. Disch, H. Kohlstedt, and D.E. B\"{u}rgler, Observation of Josephson coupling through an interlayer of antiferromagnetically ordered chromium, Phys. Rev. B \textbf{80}, 064508 (2009)
\bibitem{Zaitsev}
A. Zaitsev, G. A. Ovsyannikov, K. Y. Constantinian, Y. V. Kislinski\u{i}, A. V. Shadrin, I. V. Borisenko, and P. V. Komissinskiy, Superconducting current in hybrid structures with an antiferromagnetic interlayer, J. Exp. Theor. Phys. \textbf{110}, 336 (2010), [Z. Eksp. Teor. Fiz. \textbf{137}, 380 (2010)]
\bibitem{Zaitsev2}
G. A. Ovsyannikov, P. Komissinskiy, I. V. Borisenko, Yu. V. Kislinskii, A. V. Zaitsev, K. Y. Constantinian, and D. Winkler, Anomalous Superconducting Proximity Effect in Hybrid Oxide Heterostructures with Antiferromagnetic Layer, Phys. Rev. Lett. \textbf{99}, 017004 (2007).
\bibitem{Gorkov}
L. Gor'kov and V. Kresin, Giant magnetic effects and oscillations in antiferromagnetic Josephson weak links, Appl. Phys. Lett. \textbf{78}, 3657 (2001).
\bibitem{Andersen}
B. M. Andersen, I. V. Bobkova, P. J. Hirschfeld, and Yu. S. Barash, 0 - $\pi$ transitions in Josephson junctions with antiferromagnetic interlayers, Phys. Rev. Lett. \textbf{96}, 117005 (2006)
\bibitem{Falch}
V. Falch, J. Linder, Giant magnetoanisotropy in the Josephson effect and switching of staggered order in antiferromagnets, Phys. Rev. B \textbf{106}, 214511 (2022)
\bibitem{Eilenberger}
G. Eilenberger, Z.Phys. \textbf{214}, 195 (1968)
\bibitem{Larkin semiclass}
A. I. Larkin, and Y. N. Ovchinnikov, Quasiclassical Method in the Theory of Superconductivity, Zh. Eksp. Teor. Fiz. \textbf{55}, 2262 (1968) [Sov. Phys. JETP \textbf{28}, 1200 (1968)].
\bibitem{Fynn}
E. H. Fyhn, A. Brataas, A. Qaiumzadeh, and J. Linder, Quasiclassical theory for antiferromagnetic metals, Physical Review B \textbf{107}, 174503 (2023)
\bibitem{Aslamazov}
L. G. Aslamazov, A. I. Larkin, and Yu. N. Ovchinnikov, Josephson Effect in Superconductors Separated by a Normal Metal, Zh. Eksp.
Teor. Fiz. \textbf{55}, 323 (1968)[Sov. Phys. JETP \textbf{28}, 171 (1969)].
\bibitem{AGD}
A. A. Abrikosov, L. P. Gor'kov, and I. E. Dzyaloshinskii, Methods
of Quantum Field Theory in Statistical Physics (Dover, New York, 1975)
\bibitem{Rammer}
J. Rammer, H. Smith, Quantum field-theoretical methods in transport theory of metals, Rev. Mod. Phys. \textbf{58}, 323 (1985)
\bibitem{Reinoso}
A. Reynoso, G.Usaj, C.A. Balseiro, D. Feinberg, M.Avignon, Anomalous Josephson Current in Junctions with Spin Polarizing Quantum Point Contacts,
Phys. Rev. Lett. \textbf{101}, 107001 (2008)
\bibitem{Zazunov}
A. Zazunov, R. Egger, T. Martin, and T. Jonckheere, Anomalous Josephson Current through a Spin-Orbit Coupled Quantum Dot, Phys.Rev. Lett. \textbf{103}, 147004 (2009)
\bibitem{ISHE}
A. G. Mal'shukov, S. Sadjina, and A. Brataas, Inverse Spin Hall Effect in SNS Josephson Junctions, Phys. Rev. B \textbf{81}, 060502 (2010)
\bibitem{Liu}
 J.-F. Liu and K. Chan, Relation between symmetry breaking and the anomalous Josephson effect, Phys. Rev. B \textbf{82}, 125305 (2010)
\bibitem{Yokoyama}
 T. Yokoyama, M. Eto, Y. V. Nazarov, Anomalous Josephson effect induced by spin-orbit interaction and Zeeman effect in semiconductor nanowires, Phys. Rev. B \textbf{89}, 195407 (2014)
\bibitem{Konschelle}
 F. Konschelle,  I. V. Tokatly and F. S. Bergeret, Theory of the spin-galvanic effect and the anomalous phase shift $\varphi_0$ in superconductors and Josephson junctions with intrinsic spin-orbit coupling, Phys. Rev. B \textbf{92},125443 (2015)
\bibitem{Malsh}
A. G. Mal'shukov and C. S. Chu, Spin Hall effect in a Josephson contact, Phys. Rev. B \textbf{78}, 104503 (2008)
\bibitem{DP}
M. I. D'yakonov and V. I. Perel, Spin Orientation of Electrons Associated with the Interband Absorption of Light in Semiconductors, Zh. Eksp. Teor. Fiz. \textbf{60}, 1954 (1971) [Sov. Phys. JETP 33, 1053 (1971)].
\bibitem{Mishch}
E. G. Mishchenko, A. V. Shytov, and B. I. Halperin, Spin Current and Polarization in Impure Two-Dimensional Electron Systems with Spin-Orbit Coupling, Phys. Rev. Lett.
\textbf {93}, 226602 (2004);
\bibitem{Burkov}
A. A. Burkov, A. S. Nunez, and A. H. MacDonald, Theory of spin-charge-coupled transport in a two-dimensional electron gas with Rashba spin-orbit interactions, Phys. Rev. B \textbf{70}, 155308 (2004)
\bibitem{MalshPRLSHE}
A. G. Mal'shukov, L. Y. Wang, C. S. Chu and K. A. Chao, Spin Hall Effect on Edge Magnetization and Electric Conductance of a 2D Semiconductor Strip, Phys. Rev.Lett. \textbf {95}, 146601 (2005)
\bibitem{comment}
Some equations may differ from those in Ref.[\onlinecite{Bergeret}], because Nambu bases which are chosen in Ref.[\onlinecite{Bergeret}] and this manuscript do not coincide.
\end{thebibliography}
\end{document}